\documentclass{elsart}
\usepackage{amssymb}
\usepackage[dvips,unicode]{hyperref}
\usepackage[dvips]{graphicx}

\begin{document}

\begin{frontmatter}

\title{Flow-based analysis of Internet traffic}

\author[ST]{F. Afanasiev},
\author[SSAU]{V. Grachev},
\author[ST]{A. Petrov},
\author[SAMGAPS]{A. Sukhov\corauthref{avt}}

\address[ST]{JSV "SamaraTelecom", Aerodromnaya 45, Samara, Russia}

\address[SSAU]{Samara State Aerospace University, Moscovskoe sh. 34a,
Samara, 443086, Russia}

\address[SAMGAPS]{Laboratory of Network Technologies, Samara Academy
of Transport Engineering, 1st Bezymyanny per., 18, Samara, 443066,
Russia}

\thanks[avt]{Corresponding author\\{\em E-mail addresses:
sukhov@ssau.ru}~(Andrei Sukhov), {\em vgrachev@ssau.ru}~(Vladimir
Grachev), {\em afv@smrtlc.ru}~(Fedor Afanasiev), {\em
apetrov@smrtlc.ru}~(Anton Petrov)}

\begin{abstract}
We propose flow-based analysis to estimate quality of an Internet
connection. Using results from the queuing theory we compare two
expressions for backbone traffic that have different scopes of
applicability. A curve that shows dependence of utilization of a
link on a number of active flows in it describes different states
of the network. We propose a methodology for plotting such a curve
using data received from a Cisco router by NetFlow protocol,
determining the working area and the overloading point of the
network. Our test is an easy way to find a moment for upgrading
the backbone.
\end{abstract}

\end{frontmatter}

\section{Introduction} \label{Intr}

Modeling the traffic at the packet level has proven to be very
difficult since traffic on a link is the result of a high level of
aggregation of numerous flows. Recently, a new trend has emerged,
which consists in modeling the Internet traffic at the flow level.
A flow here is a very generic notion. It can be a TCP connection
or a UDP stream described by source and destination IP addresses,
source and destination port numbers, and the protocol number. It
is possible to get an idea about the response time of a flow and
about the distribution of the flows that are active at a certain
time in the network. From a simplicity standpoint, it is much
easier to monitor flows than to monitor packets in a router.

Network operators need to know when their backbone or peering
links have to be upgraded. As a clue for this, boundary values of
network parameters may serve - as the current values of the
network parameters reach some limit values, the links have to be
upgraded. The problem is that there is no common view point on the
set of parameters to monitor and on their boundary values. Each
provider has its own technical specifications aimed on avoiding
overloading and big providers, like Sprint \cite{FrM}, rely on the
results of their own researches. Usually, network operators
monitor peak and average link utilization levels and upgrade their
links when utilization level lays in the range 30\%-60\%.

Academic research community should play an important role in
providing analytical generalization and establishing common
terminology for processes taking place in networks. Our aim is a
paper where we accumulate our knowledge about the network quality
and related terms like length of working part, points of
overloading and etc, as well as to formulate basic recommendations
for the moments for upgrading links. Traffic measurement and
analysis should be extended to consider the case of a single path
(hop) between two routers on a high-speed backbone. This lets to
find bottlenecks on backbones like GEANT. Usually, network
operators collect statistics on packet level that include source
and destinational addresses, ports, protocols, packet flags, size,
start and end time of UDP and TCP sessions, duration of the
sessions and etc. Processing of such huge volume of data is a
difficult task demanding powerful hardware, software, big human
efforts, but, unfortunately, it does not always let to get enough
information to give some recommendations for the network under
consideration.

The contribution of the paper is a simple flow-based test to find
a moment for upgrading of a backbone. In the next Section we use
queuing theory for flow-based analysis of a backbone. Section 3
discusses three states of a network with different performances. A
test for network quality is described in the Section 4. The
results of the experiments conducted in the real networks of ISPs
are given in the Section 5.

\section{The model}
\label{model}

In this paper we model traffic as a stationary stochastic process,
using the results from the papers  Barakat et al~\cite{ChaB} and
Ben Fredj et al~\cite{BenF}.

In summary, presented model allows us to completely characterize
the data rate on a backbone link based on the following inputs:

\begin{itemize}
\item
Session arrivals in any period where the traffic intensity is
approximately constant are accurately modeled by a homogeneous
Poisson process of finite rate $\lambda$. In general, this
assumption can be relaxed to more general processes such as MAPs
(Markov Arrival Processes) \cite{Alt}, or non homogeneous Poisson
processes, but we will keep working with it for simplicity of the
analysis.

\item
The distributions of flow sizes $\{S_n\}$ and flow durations
$\{D_n\}$. In this paper we denote by $T_n$ the arrival time of
the $n$-th flow, by $S_n$ its size (e.g., in bits), and by $D_n$
its duration (e.g., in seconds).

\item
The flow rate function (shot) is $X_n(\cdot)$. A flow is called
active at time $t$ when $T_n\leqslant t\leqslant T_n+D_n$. Define
as $X_n(t-T_n)$ the rate of the $n$-th flow at time $t$ (e.g., in
bits/s), with $X_n(t-T_n)$ equal to zero for $t<T_n$ and for
$t>T_n+D_n$. In other words, $X_n(t-T_n)$ is zero if flow $n$ is
not active at time $t$.

\end{itemize}

Define $R(t)$ as the total rate of data (e.g., in bits/s) on the
modeled link at time $t$. It is the result of the addition of the
rates of the different flows. We can then write
\begin{equation}
\label{totr} R(t)=\sum_{n\in \mathbb Z} X_n(t-T_n)
\end{equation}
The process from Eq.~(\ref{totr}) can describe the number of
active flows found at time $t$ in an $M/G/\infty$ queue
\cite{kler}, if $X_n(t-T_n)=1$ at $t\in [T_n,T_n+D_n]$.

The model presented by Barakat et al \cite{ChaB} is able to
compute the average and the variation of the backbone traffic. In
summary:
\begin{itemize}
\item
The average total rate of the traffic is given by the two
parameters $\lambda$ and ${\mathbb E}[S_n]$:
\begin{equation}
\label{srR} {\mathbb E}[R(t)]=\lambda {\mathbb E} [S_n]
\end{equation}

\item
The variance of the total rate $V_R$ (i.e., burstiness of the
traffic) is given by the two parameters $\lambda$ and ${\mathbb
E}[S_n^2/D_n]$:
\begin{equation}
\label{srVR} V_R=\lambda {\mathbb E} [S_n^2/D_n]
\end{equation}

\end{itemize}

It should be mentioned that Eq.~(\ref{srR}) is true only for the
ideal case of the backbone link of unrestricted capacity that can
be applied to underloaded links. The arrival rate of flows
$\lambda$ describes the user's behaviour that doesn't depend on
the network utilization. The cumulative number of flows that
arrive at a link will remain linear even if the network doesn't
satisfy all the incoming demands. The average flow size does not
also depend on a specific system, but only on the current
distribution of documents sizes found in the Internet. Thus the
main drawback of the ratio~(\ref{srR}) is lack of its definite
usage limits, which is the result of the fact that variables
describing the system are in no way connected with its current
state.

Ben Fredj et al~\cite{BenF} consider a case where the link
capacity $C$ is very large compared to the external rate limits
and such that it is virtually transparent. By this they mean that
the probability that the sum of external rate limits of all active
flows exceeds link capacity $C$ is negligibly small. This
assumption is reasonable for the large, moderately loaded links of
major backbone providers.

The number of active flows is now unconstrained by the considered
link which appears as an $M/G/\infty$ queue. Let ${\mathbb
E}[D_n]$ be the mean duration of flow and $N$ the mean number of
active flows. By Little's law we have:
\begin{equation}
\label{srN} N=\lambda {\mathbb E} [D_n]
\end{equation}
This formula describes the network state more precisely as the
average number of active flows on the bandwidth unit increases
with the utilization. In other words, the average duration of flow
enables us to judge the real network state in contrast to its
average size.

\section{Performance states on flow level}
\label{perf}

Analyzing the Eqs.~(\ref{srR}) and (\ref{srN}) we may compare the
ideal and the real states of the network link under consideration.
In order to analyze the quality of the backbone area or link to
provider we are going to construct a graphical dependence between
the link loading and the number of active flows in it.

\begin{figure}
\begin{center}
\includegraphics[width=0.85\textwidth]{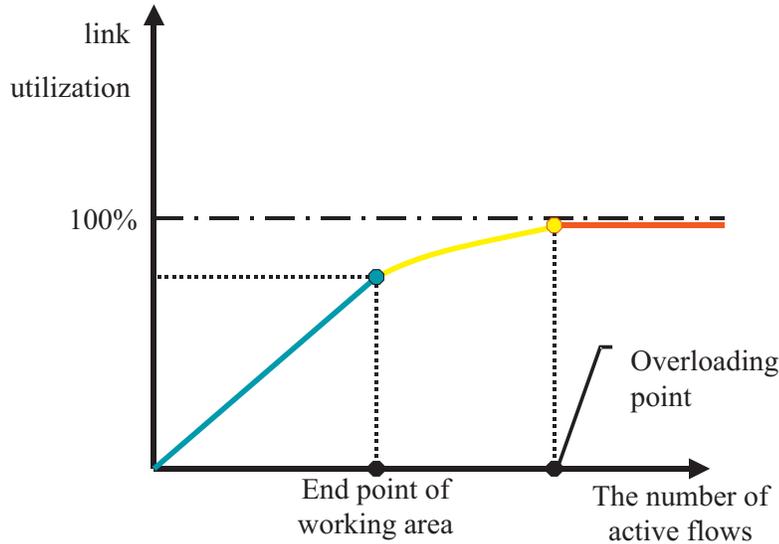}
\end{center}
  \caption{Link utilization vs the number of active flows}
  \label{nagr}
\end{figure}

The suggested curve has been found on Fig.~\ref{nagr}. On the
given curve three parts corresponding to the different network
states can be identified.

The first part of the curve highlighted in green describes the
network state close to the ideal, which coincides with the domain
of applicability of the more strict state equation~(\ref{srR}).
The possibility of an error considered, the points corresponding
to the real network states are supposed to be on the straight
line, which corresponds to the working network area. The end of
the straight line, which can be found from an experiment, defines
thus the length of the working area.

The second part of the curve highlighted in yellow corresponds to
the moderately loaded network. This parts coincides with the scope
of less strict equation of state~(\ref{srN}), when the diversion
from the ideal network state becomes obvious. Increased average
duration of a flow compared to the working area and therefore a
larger number of active flows on the bandwidth unit are
characteristic of this network state.

The red part of the curve corresponds to the totally disabled
network with considerable loss of packets.

\section{Test for network quality}
\label{test}

In order to prove our hypothesis we made experiment on border
gateway routers of Ireland National Research and Education Network
- HEANet and Russian ISP "SamaraTelecom". Both networks have
several internal and external links but basic load lies on one
channel to Global Internet (622 Mbps for HEANet, 8 Mbps for ST).
The utilization of these links varies in wide limits from 5\% to
90\% with a clearly identifiable busy period.

A passive monitoring system based on Cisco's NetFlow \cite{Cis}
technology was used to collect values of links utilization and
number of active flows in real-time.

We measured on the Cisco 7206 router with NetFlow switched on. The
detailed description of Cisco NetFlow can be found in
documentation of Cisco \cite{Cis}.

\begin{figure}
\begin{center}
\includegraphics[width=0.45\textwidth]{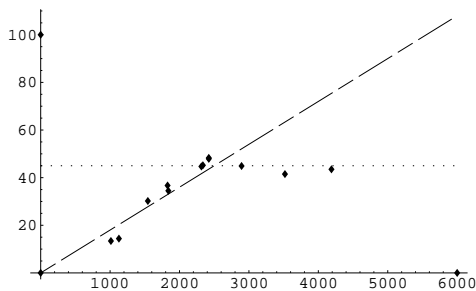}
\end{center}
  \caption{ Link $U$  utilization vs the number of active flows
  $N_a$  for Samara Telecom network}
\label{zav}
\end{figure}

It should be mentioned that exporting of flows is to be switched
on all interfaces from which the data is to be collected.
Otherwise the data obtained while measuring won't contain complete
information. For example, we didn't quite succeed in our
experiment with Ireland National Research and Education Network
(HeaNet) because of the incomplete setup of NetFlow support on the
routers. Usually a network has only one or a few external links
while internal connections remain considerably less loaded. That's
why we can construct dependence of the number of all active flows
on the boundary router on the general loading of the external
links. It is achieved by the following commands:

\begin{itemize}
\item
{\sl sh ip cache flow} - gives information about the number of
active and inactive flows, about the parameters of the flows at
the real time.
\item
{\sl sh int} {\sf [a name for the external interface]} - gives
information about the current link loading.
\end{itemize}

The data obtained with the commands contain all the necessary
values for the construction of the curve similar to the one drawn
in Fig.~\ref{nagr}. The values are to be recorded twenty-four
hours a day every 30 minutes during a week to discover network
behaviour while differently loaded. It's quite easy to write a
script, which will collect the data from the router to the
management server.

\section{Experimental validation}
\label{exper}

It has already been mentioned that such a test has been conducted
on the boundary router of Samara Telecom Company. As external
links four $E1$ links (4 x 2.048 Mbit/s) to different ISP were
used. We have taken a number of points for different network
loading states. The results of this measurement are represented on
Fig.~\ref{zav}. On the $X$ axis there is the number of flows and
on the $Y$ axis there is the network loading in percents from its
maximal value.

The separate points on the curve from Fig.~\ref{zav} correspond to
the real network states and the discrete straight line, which
consists of dashes of different length, depicts the ideal network
behaviour corresponding to the Eq.~(\ref{srR}). The angle of of
this straight line is found as average of $U/N_a$ ratio taken from
the states when the network loading didn't exceed $40\%$.

The working area is limited from the top by a straight dotted
line, which is the nearest approach to the description of the
network behaviour when the loading is heavy. The crossing of the
two straight lines allows finding the length of the working area
$(45\%)$.  When the number of flows is more than 2500 network
experiences overloading which leads to the slowing down of the
speed of the network connections. The overall link loading
practically doesn't increase along with the amount of requests and
the connection quality becomes almost twice as bad. It should be
mentioned that in the network under consideration transition area
or moderately loaded network doesn't virtually exist. (see
Fig.~\ref{nagr}). The experiment conducted is evidence of the fact
that providers do not always guarantee the work within the working
network area which is especially important for the applications
requiring high network quality, for example, IP telephony and
videoconferencing. At the hours when the network loading was the
heaviest the actual network capacity proved to be nearly twice as
low as stated.

\section{Feature work}
\label{feature}

In this paper some methods are described which allow us to
evaluate Internet link quality on the base of flow technology and
increase their capacity in time. At the moment we are working on
developing utilities, which will make it possible to construct
dependence of link loading on the number of active flows
automatically and calculate the length of the working area.

We also plan further research work, which suggests evaluation of
the flow parameters and above different speed of data transfer. We
propose to derive analytic dependence of flow speed, i. e. to
evaluate the time of transfer of file of a certain size between
the two remote IP addresses (end-to-end) on the packet delivery
time and percentage loss. In other words to theoretically evaluate
connection quality according to ping data and its analogue for TCP
packets.

Unfortunately, low-speed Russian Internet links are only available
for our research work. Our attempts to collect the necessary data
from high-speed international links with preferable speeds of 622
Mbps and 2.4 Gbps failed. As we know from our own experience one
can get the data in a necessary format only through personal
communication with the staff of the researched network. That's why
we need assistance and joint projects for financing the trips to
the place of research work.


\begin{thebibliography}{00}

\bibitem{Alt}
E. Altman, K. Avratchenkov, and C. Barakat, A stochastic model for
TCP/IP with stationary random losses, ACM SIGCOMM, September 2000.

\bibitem{ChaB}
Chadi Barakat, Patrick Thiran, Gianluca Iannaccone, Christophe
Diot, Phillipe Owezarski, A flow-based model for Internet backbone
traffic, IMW 2002 (Internet Measurement Workshop)

\bibitem{BenF}
S. Ben Fredj, T. Bonald, A. Proutiere, G. Regnie and J. Roberts,
Statistical Bandwidth Sharing: A Study of Congestion at Flow
Level, ACM SIGCOMM, August 2001.

\bibitem{BrM}
N. Brownlee, C. Mills, G. Ruth, Traffic Flow Measurement:
Architecture (RFC 2722), October 1999

\bibitem{Cis}
Cisco IOS NetFlow site, Cisco Systems,
http://www.cisco.com/go/netflow/

\bibitem{FrM}
C. Fraleigh, S. Moon, C. Diot, B. Lyles and F. Tobagi,
Packet-Level Traffic Measurements from a Tier-1 IP Backbone,
Sprint ATL Technical Report TR01-ATL-110101, November 2001

\bibitem{kler}
L. Kleinrock, Queueing Systems, Vol. I: Theory, Wiley, 1975.

\bibitem{JitP}
Jitendra Padhye, Victor Firoiu, Don Towsley, Jim Kurose, Modeling
TCP Throughput: A Simple Model and its Empirical Validation, ACM
SIGCOMM, September 1998.


\end{thebibliography}
\end{document}